\documentclass{article}
\usepackage{LaThuileFPSpro}
\begin{document}
\title{ 
Implications of right-handed neutrinos with electroweak-scale masses}
\author{
  P. Q. Hung      \\
  {\em Dept. of Physics, University of Virginia,
382 McCormick Road, P. O. Box 400714,} \\
{\em Charlottesville, Virginia 22904-4714, 
USA} 
}
  
\maketitle

\baselineskip=11.6pt

\begin{abstract}
The possibility of constructing a model in which right-handed neutrinos have
electroweak-scale masses as well as being {\em non-sterile} was espoused in
\cite{hung1}.In this talk, I will review the ideas and results of \cite{hung1}
and discuss its implications for colliders such as the Tevatron, LHC and ILC.
\end{abstract}
\newpage
\section{Introduction}
The origin of neutrino masses and the puzzle over their smallness are
two of the outstanding questions in particle physics. Of related importance
is the nature of the neutrinos: Are they Dirac or Majorana particles? 
There is no doubt about the importance that neutrinos have in particle 
physics and cosmology: The understanding of their masses unquestionably points
to features that cannot be explained by staying strictly within the Standard
Model (SM) such as, for example, the baryon number asymmetry which
might arise through the so-called leptogenesis coming from the decay
of a heavy Majorana neutrino.
Furthermore, results from neutrino oscillation data indicated
a mixing matrix in the lepton sector which is markedly different from
that coming out of the quark sector. One cannot help but wonder if,
despite this dissimilarity, the two sectors can ``learn'' from each
other. 

Neutrino masses are believed to be be tiny compared with other fermion
masses, of order O($<1\,eV$). Why this is so is one of the biggest
mysteries which we are trying to unlock. The ``simplest'' way to give
a mass to the neutrino is to add a SM singlet right-handed neutrino
to the SM and give it a Dirac mass. However, to account for the
smallness of the neutrino masses, Yukawa couplings of O($10^{-11}$) have
to be put in {\em by hand}. This is generally considered to be unnatural
unless there are dynamical or symmetrical reasons for it to be so
\cite{dirac}. The by-far most popular scenario is the quintessential
see-saw mechanism \cite{seesaw} where, in addition to the Dirac mass 
($m_D$) term which
couples left- and right-handed neutrinos, a lepton-number-violating
Majorana mass ($M_R$) term for the right-handed (the simplest version) 
neutrinos is written down. In the ``standard'' see-saw mechanism,
this Majorana mass term is {\em huge} compared with the Dirac mass
term (which is proportional to the electroweak scale) resulting in a 
tiny mass $\sim m_D^2/M_R$ for the lighter of
the two eigenstates. The right-handed neutrinos being sterile
in this scenario and being extremely heavy are practically undetectable,
at least directly. Therefore, in its simplest version, one just cannot
directly verify the see-saw mechanism since one cannot detect the
right-handed neutrinos. Are there other ways? 

Since, within the framework of see-saw scenarios,
the light neutrino sector is only sensitive to the ratio
$m_D^2/M_R$, it is legitimate to ask how one can change $m_D$ and
$M_R$ in such a way as to keep $m_D^2/M_R$ unchanged. The question is
the following: Could one lower both of them in such a way as to make
$M_R$ slide into a region, in particular around the electroweak scale, 
where one could have an access to the right-handed neutrino sector? 
This is one of the motivations for the construction of a model
of electroweak-scale right-handed neutrino mass \cite{hung1}. The organization
of the talk will be as follows. First, a brief review of the see-saw mechanism
will be presented. Next, we will present arguments on why the right-handed
neutrinos can be as light as or lighter than the electroweak scale. We then
discuss the implications of electroweak-scale $\nu_R$'s, including 
the production and decays of $\nu_R$'s as well as lepton-number
violating processes at colliders. A conclusion will follow the phenomenological
discussion.

\section{The see-saw mechanism in a nutshell}

In the ``standard'' see-saw scenarios \cite{seesaw}, $\nu_R$'s are SM {\em singlets} and
are commonly termed {\em sterile}. This has obviously deep implications
on the nature and sizes of the Dirac and Majorana masses.
\begin{itemize}

\item Dirac Mass:

The neutrino Yukawa interaction with a sterile right-handed neutrino 
which gives rise to the Dirac mass term is usually written as
\begin{equation}
\label{dirac}
{\cal L}_D = g_L\,\bar{l}_L \phi \, \nu_R + H.c. \,,
\end{equation}
where $l_L = (\nu_L, e_L)$ and $\phi = (\phi^0, \phi^-)$ are 
the usual SM $SU(2)_L$ doublets. 
When $\langle \phi \rangle = (\Lambda_{EW}/\sqrt{2}, 0)$
with $\Lambda_{EW} \approx 246\, GeV$, one obtains the following the neutrino Dirac mass
\begin{equation}
\label{md}
m_D = g_L \,\Lambda_{EW}/\sqrt{2} \,.
\end{equation}
In consequence, the Dirac mass is proportional to the electroweak scale 
$\Lambda_{EW}$, although it crucially depends on an arbitrary Yukawa coupling
$g_L$. It is worth to emphasize again that this is the case because $\nu_R$ is
a SM singlet. We will see below that when $\nu_R$ is {\em not} a SM singlet,
the Dirac mass will no longer be related to $\Lambda_{EW}$.

\item Majorana mass:

The source of the right-handed neutrino Majorana mass is quite model-dependent,
although it is commonly found within the framework of a Grand Unified Theory (GUT).
In what follows, we will write it simply as
\begin{equation}
\label{maj} 
{\cal L}_M = M_R \,\nu_R^{T}\,\sigma_2\,\nu_R \,.
\end{equation}
The above Majorana mass term violates lepton number by two units.

\item Mass eigenvalues:

The two well-known eigenvalues are $\sim -m_D^2/M_R$ and $M_R$
for $M_R \gg m_D$. The two neutrino mass eigenstates
which are now Majorana particles are approximately the left-handed
neutrino for the lighter state and the right-handed neutrino for the
heavy state. Since $m_D$
is proportional to the electroweak scale (modulo the unknown Yukawa coupling),
a light neutrino with mass of order O($<1\,eV$) in general requires $M_R$
to be of order O($\sim 10^{13}\,GeV$). In this type of scenarios, one just
{\em cannot directly} detect the right-handed neutrinos.

Since neutrinos (both the light state and the heavy state) are now
Majorana particles, it is therefore of utmost importance to test
this feature of the model. One should look for processes that violate lepton
number conservation. In the light sector, one could look for
neutrinoless double beta decay for example which gives an upper bound,
not on the mass of the light state, but on the combination
$<m_{\beta\,\beta}> = [\sum |U_{ei}|^2 m_{i}^2]^{1/2} < 0.35\,eV$, where
$m_i$ are the light masses \cite{0nubeta}. This search is not easy
because of various nuclear details. This is where the
right-handed neutrino sector comes in if the right-handed
neutrinos are light enough.
As for the heavy sector, at least in its
simplest version, there is no such a possibility for testing the Majorana
nature of the right-handed neutrinos. Electroweak-scale SM {\em singlet}
right-handed neutrinos were contemplated as a possibility which could
enable one to probe the right-handed sector. There are however a number
of delicate issues with these scenarios which might
prevent its observability unless some fine tuning is realized \cite{smirnov}.
An extensive list of references of works dealing with ``light'' right-handed
neutrinos can be found in \cite{smirnov}.

\end{itemize}

Can the right-handed neutrinos be {\em non-sterile}? If one can
construct such a scenario then one can hope to be able to find
them at colliders and test the Majorana nature of neutrinos.
In what follows I will describe a model in which right-handed neutrinos
are both ``light'', i.e. having electroweak-scale masses, and ``observable'',
i.e. transforming non-trivially under the SM gauge group.

\section{A Model of electrowek-scale right-handed neutrino mass}

The objective of \cite{hung1} was to construct a model in which $\nu_R$'s
are {\em not} sterile and have a {\em low} mass of O($\Lambda_{EW}$).
There are two constraints that have to be satisfied in the construction
of such a model.
\begin{itemize}

\item A non-sterile $\nu_R$ will couple to the
Z boson. There is however a strong constraint from
the Z width: There are only {\em three} light left-handed
neutrinos.

\item A Majorana bilinear $\nu_R^{T}\,\sigma_2\,\nu_R$
will transform {\em non-trivially} under
$SU(2)_L \otimes U(1)_Y$. This imposes a strong constraint on the Higgs field 
which couples to that bilinear and
which develops a non-zero vacuum expectation value, namely one
has to preserve the successful relation 
$M_W = M_Z\,\cos \theta_W$!

\end{itemize}
As we shall see below, the first constraint sets a {\em lower bound} on the
right-handed neutrino mass while the second will determine the enlargement
of the Higgs structure of an extended SM.

The simplest possibility and the one that was used in \cite{hung1} is to
put $\nu_R$ into a doublet of $SU(2)_L$. If it belongs to a doublet then
its partner would be a {\em negatively} charged right-handed lepton. 
Could it be the right-handed SM charged lepton? The answer is negative
because neutral current experiments have shown that the SM right-handed charged
leptons are $SU(2)_L$ singlets. In consequence, this right-handed
charged lepton has to be a new type: the so-called {\em mirror} lepton.
We write this new doublet as follows
\begin{equation}
\label{mirror}
l^{M}_R = \left( \begin{array}{c}
\nu_R \\
e^{M}_{R}
\end{array} \right) \,,
\end{equation}
where now the left-handed charged mirror lepton, namely $e^{M}_{L}$, is
a SM singlet. So, the above doublet plus $e^{M}_{L}$ will be the mirror
copy of the SM doublet $l_L = (\nu_L, e_L)$ and $e_R$.

In a similar fashion to the ``standard'' see-saw scenario, one can
write down the interactions which will give a Dirac mass term
for the neutrino and a Majorana mass term.
\begin{itemize}

\item Dirac mass:

A Dirac mass term for the neutrino is proportional to
$\bar{l}_L\,l^{M}_R$. This combination can couple to a SM
{\em singlet} scalar field $\phi_S$ as follows
\begin{equation}
\label{diracmass}
{\cal L}_S = g_{Sl} \, \bar{l}_{L}\, \phi_S \, l^{M}_{R} + H.c \,.
\end{equation}
 When $\phi_S$ develops a non-vanishing VEV, namely $\langle \phi_S \rangle = v_S$,
the neutrino Dirac mass takes the form
\begin{equation}
\label{diracmass2}
m_D = g_{Sl}\,v_S \,.
\end{equation} 
In this model, the Dirac mass is {\em not} linked to the electroweak scale. We will
see below the range of values that $v_S$ can take.

Notice that for the charged fermions (leptons and quarks), there are additional
couplings to $\phi_S$ involving the $SU(2)_L$ singlets of the 
forms $\bar{f}^{M}_L\,f_R$, where $f$ stands for $q$ or $e$. For simplicity,
one can assume similar Yukawa couplings to the ones given in the above form.
This yields the mixing given in \cite{hung1}. There it was shown that
the mixing between SM and mirror charged fermions is {\em negligible}.

\item Majorana mass:

In our model, the lepton-number violating relevant fermion bilinear is 
$l^{M,T}_R \, \sigma_2 l^{M}_R$. This transforms as a singlet and as a triplet
of $SU(2)_L$. A singlet Higgs field which couples to this bilinear and which
develops a VEV would break charge conservation. The only other option
is a triplet Higgs $\tilde{\chi}=(3,Y/2=1)$ which is written explicitely as
\begin{equation}
\label{triplet}
\tilde{\chi} = \frac{1}{\sqrt{2}}\,\vec{\tau}.\vec{\chi}=
\left( \begin{array}{cc}
\frac{1}{\sqrt{2}}\,\chi^{+} & \chi^{++} \\
\chi^{0} & -\frac{1}{\sqrt{2}}\,\chi^{+}
\end{array} \right) \,.
\end{equation}
which couples to the bilinear as follows
\begin{equation}
\label{majorana}
{\cal L}_M = g_M \,l^{M,T}_{R}\, \sigma_2 \,\tau_2 \,
\tilde{\chi}\, l^{M}_{R} \,.
\end{equation}
With $\langle \chi^{0} \rangle = v_M$, the Majorana mass is now
\begin{equation}
\label{majorana2}
M_R = g_M\,v_M \,.
\end{equation}
The above VEV breaks $SU(2)_L$. The successful relation 
$M_W = M_Z\,\cos \theta_W$ ($\rho=1$ at tree level)
which relies primarily on $SU(2)_L$ Higgs fields being
doublets would be spoiled unless $v_M \ll \Lambda_{EW}$. This
is a severe constraint that needs to be addressed in our model.

An important remark is in order here. In order to prevent the
left-handed neutrinos to acquire a Majorana mass of the same order
asthe right-handed one as well as to prevent a large Dirac mass
(coupling of $\bar{l}_L\,l^{M}_R$ to $\tilde{\chi}$), a
global $U(1)_M$ symmetry is imposed \cite{hung1} (and explicitely
broken by the Higgs sector). A tiny Majorana mass for the left-handed
neutrinos arises at one-loop level \cite{hung1}.

An elegant solution to this problem was provided about twenty
years ago by \cite{custodial}: If the Higgs potential which now
includes triplet scalars possesses a custodial symmetry such that
$M_W = M_Z\,\cos \theta_W$ is preserved at tree-level then the
triplet VEV's can be as large as the electroweak scale.
$\rho=1$ is therefore the manifestation of 
an approximate {\em custodial}
global $SU(2)$ symmetry of the Higgs potential. To
maintain that {\em custodial symmetry}, one can add an additional
Higgs triplet $\xi=(3,Y/2=0)$ which can be grouped with
$\tilde{\chi}=(3,Y/2=1)$ to form
\begin{equation}
\label{chi}
\chi = \left( \begin{array}{ccc}
\chi^{0} &\xi^{+}& \chi^{++} \\
\chi^{-} &\xi^{0}&\chi^{+} \\
\chi^{--}&\xi^{-}& \chi^{0*}
\end{array} \right) \,,
\end{equation}
where the full potential now exhibits a global $SU(2)_L \otimes SU(2)_R$
symmetry. The following VEV of $\chi$ breaks $SU(2)_L \otimes SU(2)_R$
down to a custodial $SU(2)$ symmetry
\begin{equation}
\label{chivev}
\langle \chi \rangle = \left( \begin{array}{ccc}
v_M &0&0 \\
0&v_M&0 \\
0&0&v_M
\end{array} \right) \,.
\end{equation}
This gives
\begin{equation}
\label{wzmass}
M_W = g\,v/2 \,;\, M_Z = M_W/\cos \theta_W \,,
\end{equation}
with
\begin{equation}
\label{vevvalue}
v= \sqrt{v_2^2 + 8\,v_M^2} \,,
\end{equation}
and
\begin{equation}
\label{vevvalue2}
\langle \Phi \rangle = v_2/\sqrt{2} \,,
\end{equation}
where $\Phi$ is a doublet. The nice feature of this scenario is
the fact that now $v_M$ {\em can} be of the order of the electroweak scale
{\em without} spoiling $\rho=1$. As discussed in \cite{hung1}, there are
no massless NG bosons in this model since $U(1)_M$ is explicitely broken.

The upshot of all this is the following nice result
\begin{equation}
\label{mr}
M_R \sim O(\Lambda_{EW}) \,.
\end{equation}
The right-handed neutrino mass can now be {\em naturally} of the order of the
electroweak scale (but not more)!

How low can $M_R$ be? A right-handed neutrino with a mass lower than half
the Z-boson mass would contribute to the Z width with the amount as the
left-handed one. This is ruled out experimentally. We therefore conclude
that $M_R$ lies in a rather ``narrow'' range
\begin{equation}
\label{mrbound}
M_Z/2< M_R < \Lambda_{EW} \,.
\end{equation}

\item Estimate of the singlet Higgs VEV:

With the light neutrino mass $m_{\nu} \leq 1\,eV$ and $M_R \sim O(\Lambda_{EW})$,
one can get a rough estimate on the singlet VEV by putting
$g_{SL} \sim O(1)$ to give
\begin{equation}
\label{singletvev}
m_D \sim v_S \sim 10^{5}\,eV \,.
\end{equation}
A small scale such as $v_S$ is interesting in many respects. First there appears
to be some kind of hierarchy problem since $v_S$ is six orders of magnitude
smaller than $v_M$, although it is not as severe as the GUT hierarchy problem.
However, one can imagine that $v_S$ might actually be the {\em present}
classical value of the singlet Higgs field $\phi_{S}(t_0)$ whose effective
potential might be of a ``slow-rolling'' type. This type of scenario
was proposed in a mass-varying neutrino (MaVan) model of the first reference
of \cite{mavan}. The Dirac will keep increasing until $\phi_{S}$ reaches
the true minimum which could be of the order of the electroweak scale itself! 

\end{itemize}

What (\ref{mr}) and (\ref{singletvev}) tell us is that, in our scenario, 
the mass scales participating in the see-saw mechanism are slided ``downward''
with respect to the ``standard'' see-saw scenario, but now there is one 
phenomenological advantage: One can now search for the right-handed neutrinos
at colliders. As we have mentioned above, the light neutrinos are only
sensitive to the ratios $m_D^2/M_R$ and not directly to the scale $m_D$.
A discovery of an electroweak-scale right-handed neutrino would greatly help
us determine what $m_D$ should be. We now turn to the discussion on
the detectability of the electroweak-scale right-handed neutrinos.

\section{Phenomenology of Electroweak Scale $\nu_R$'s}

Since we are dealing with {\em Majorana neutrinos}
wirh electroweak scale masses, it is not surprising that we should
expect lepton-number violating processes at electroweak scale energies. 
In particular, we should be able to produce $\nu_R$'s
and observe their decays at colliders (LHC, etc...).
The characteristic signatures will be 
{\em like-sign dilepton} events which are a high-energy equivalent
of neutrinoless double beta decay.

Since $\nu_R$'s are members of $SU(2)_L$ doublets 
$l^{M}_R = \left( \begin{array}{c}
\nu_R \\
e^{M}_{R}
\end{array} \right)$, they interact with the Z and W bosons. They
are {\em no longer} sterile! Let us now recall that we have
the constraint $M_Z/2< M_R < \Lambda_{EW}$. This means that, in
principle, $\nu_R$'s can be produced at colliders, being sufficiently
light. Unlike the case with low-mass singlet $\nu_R$'s whose
production at colliders could be suppressed, the right-handed neutrinos 
in our scenario couple directly to the Z boson and the production
of a pair of $\nu_R$'s is unsuppressed. One has
\begin{equation}
\label{production}
q + \bar{q} \rightarrow Z
\rightarrow  \nu_R + \nu_R \,.
\end{equation}
Since $\nu_R$'s are Majorana particles, they can have transitions
such as $\nu_R \rightarrow l^{M,\mp}_R + W^{\pm}$.
A heavier $\nu_R$ can decay into a lighter $l^{M}_R$ and one can have
\begin{equation}
\label{production2}
\nu_R + \nu_R \rightarrow l^{M,\mp}_R +l^{M,\mp}_R +W^{\pm} + W^{\pm}
\rightarrow l^{\mp}_L +l^{\mp}_L +W^{\pm} + W^{\pm}
+ \phi_S + \phi_S \,,
\end{equation}
where $\phi_S$ would be missing energy. This gives rise to
interesting {\em like-sign} dilepton events. Since this involves
missing energy, one would have to be careful with background.
For example,
one of such backgound could be a production of
$W^{\pm}\,W^{\pm}\,W^{\mp}\,W^{\mp}$ with 2 like-sign
W's decaying into a charged lepton plus a neutrino (``missing energy'').
But...This is of O($\alpha_{W}^2$) in amplitude smaller than the
above process. In addition, depending on the lifetime of the
mirror leptons, the SM leptons appear at a displaced vertex.
Lepton-number violating process with like-sign dileptons
can also occur with $\nu_R$'s in the intermediate
state (from $W^{\pm}\, W^{\pm} \rightarrow 
l^{\pm}_L +l^{\pm}_L$) but that involves very small mixing angles
of the order $\frac{m_{\nu}}{M_R}$.

In consequence, within the framework of our model, one has the
interesting prospect of producing and detecting electroweak-scale
right-handed neutrinos through lepton-number violating processes
such as like-sign dileptons as described above. Detailed
phenomenological analyses are in progress.

\section{Other phenomenological consequences}

There are several other interesting consequences of the model which are
currently under investigation. One of such consequences involves the
phenomenology of the triplet Higgses that exist in this model:
$\tilde{\chi}$ and $\xi$. Since they carry electroweak quantum numbers,
they can be produced at colliders such as the LHC or ILC.
The various scalars in $\tilde{\chi}$ couple 
to the mirror fermions through Eq. (\ref{majorana}) and 
can possibly searched for through the decays of the mirror fermions.
$\xi$ does not couple directly to fermions (SM and mirror) and
the various components would decay either directly to a pair
of electroweak gauge bosons either real or virtual.

The mirror fermions carry exactly the same quantum numbers as the SM
fermions. They can be produced in exactly the same manner as
the SM fermions at colliders. However their decays will be quite
interesting. From Eq. (\ref{diracmass}), one can see that
a charged mirror fermion can decay into its SM counterpart plus
$\phi_S$ which would be missing energy. This kind of decay for the
charged mirror leptons has already been mentioned above 
(\ref{production2}).

Last but not least, vacuum stability considerations will link
the masses of the scalar sector which now includes the triplet
Higgses to those of the fermions (SM and mirror). This is under
preparation.

\section{Conclusion}

\begin{itemize}

\item It is possible to have a seesaw mechanism in which the Majorana
mass of the right-handed neutrinos can be of the order of the
electroweak scale and, in fact, can be situated in
a ``narrow'' range $M_Z/2< M_R < \Lambda_{EW}$. There is {\em no}
reason why it should be close to some GUT scale.

\item The lepton-number violating processes coming from the ``heavy''
{\em non-sterile} $\nu_R$'s can now be accessible
{\em expermentally} at colliders! In contrast, in models where
$\nu_R$'s are SM singlets, it is problematic to both have
a light neutrino and a non-negligible coupling between sterile
and active neutrinos, resulting in a situation in which it
might be extremely hard to detect lepton-number violating
processes at the LHC for example \cite{smirnov}.

\item There is a rich spectrum of particles which can be tested in a 
not-too-distant future.

\end{itemize}

Below is a grossly incomplete list of references. My apologies for
not being able to list all the references because of length restrictions.
%
%
\section{Acknowledgements}
I would like to thank Mario Greco and the organizers of La Thuile 07
for an exciting conference and the Aspen Center for Physics where part
of this manuscript is written. This work is supported
in parts by the US Department of Energy under grant No.
DE-A505-89ER40518.
%

%
\end{document}